# Evaluating Personal Archiving Strategies for Internet-based Information

*Catherine C. Marshall; Microsoft; San Francisco, CA; Frank McCown and Michael L. Nelson; Old Dominion University; Norfolk, VA*

## Abstract

*Internet-based personal digital belongings present different vulnerabilities than locally stored materials. We use responses to a survey of people who have recovered lost websites, in combination with supplementary interviews, to paint a fuller picture of current curatorial strategies and practices. We examine the types of personal, topical, and commercial websites that respondents have lost and the reasons they have lost this potentially valuable material. We further explore what they have tried to recover and how the loss influences their subsequent practices. We found that curation of personal digital materials in online stores bears some striking similarities to the curation of similar materials stored locally in that study participants continue to archive personal assets by relying on a combination of benign neglect, sporadic backups, and unsystematic file replication. However, we have also identified issues specific to Internet-based material: how risk is spread by distributing the files among multiple servers and services; the circular reasoning participants use when they discuss the safety of their digital assets; and the types of online material that are particularly vulnerable to loss. The study reveals ways in which expectations of permanence and notification are violated and situations in which benign neglect has far greater consequences for the long-term fate of important digital assets.*

## Introduction

Our past work has revealed that people archive their personal digital belongings by relying on a combination of benign neglect, sporadic backups, and unsystematic file replication. Even the most valuable of their digital assets – files representing considerable investments of effort, significant emotional worth, or actual cash expenditures – are often in danger of being lost. Distributed storage, uncontrolled accumulation of digital materials, a lack of standard curation practices, and an absence of long term retrieval capabilities all point toward an incipient digital dark age [1,2].

But does this threat of unintentional impermanence hold for personal assets stored on network services? Research on lazy preservation techniques – ways of recovering content from large scale digital holding pens such as the Internet Archive and Google's extensive cache – suggests that Internet-based information presents different vulnerabilities than content stored locally on personal computers and home networks [3,4].

The types of material that people publish and store on the Internet may be different than the files they store and manage locally; this material may change less frequently, be less private, or be explicitly published for other people to read and access. Furthermore, individuals may have less control over network-based storage: they may be buying this storage from a service provider or using server space maintained by an institution such as a university. Network-based storage may also offer (or be perceived as offering) different safety nets than local storage. Even the material that is not shared – web-based email, personal photo stores, digital briefcases, and other non-published files – may be perceived differently, as less at risk, than local files.

Consumers recognize the distinctions between materials stored locally and materials stored on the Internet. For example, people may express the related opinion that if you find something on the Internet once, it will be there when you look for it again, suggesting an almost magical persistence (in fact, one participant in the study described in [1] said, "I thought that they [web pages] were all set in stone."). While individuals usually attribute this characteristic to files they have found rather than to personal material that they have stored on Internet services, there is no in-principal reason that this belief would not extend to their own digital belongings. Moreover, consumers trust services such as Yahoo Mail and Flickr – and sometimes their own Internet Service Providers (ISPs) – to such a great extent that they may believe that it is unnecessary to safeguard the associated content themselves. In short, Internet-based material is considered safe in a way that local files are not.

In this paper, we investigate this perception and its consequences for personal digital archiving using a detailed survey and follow-on interviews of people who have attempted to recover their files through Warrick, a utility that rebuilds websites using large-scale Internet caches and archives [3], or directly from the Internet Archive's Wayback Machine [5]. The interviews are oriented toward discovering broad motivations and values – the "whys" – while the survey helps us characterize what happened in specific instances of website crash and recovery. In essence, we investigate how and whether individuals protect their server- and service-based files, how they lose important material, what their impulse is when they go about trying to recover it, and why it is important to them to get this content back.

Since a website represents a slice of a person's digital belongings, we will first characterize what kind of creative effort is represented by the missing websites, and what types of files have been lost. Although some of the material is personal (for example, a portfolio of past work), other portions are contributed through social interaction (for example, via forums, multi-author blogs, or wikis associated with a topical or fan website). Still other content accumulates over time (for example, a database associated with a small-scale e-commerce website). A website may also involve a broad range of file types – text, images, audio, video, formatted documents, code implementing dynamic or interactive capabilities, and other common forms – that will help us understand more general problems.

Once we have characterized what kinds of material constitute these personal websites and which material was sufficiently valuable to seek its return, we will explore the failed personal archiving strategies that led to the loss. Because some of the materials were maintained on behalf of others, we will also be able to exam-ine the broader problems of the *ad hoc* IT practices characteristic of home and small business users. We will also examine the ways in which respondents lost their web-based digital belongings, how they discovered the loss, and whether this loss (and potential recovery) has changed their behavior at all. Finally, we reflect on what these findings imply for personal digital archiving.

## Study Description

This study combines two different data sources: a self-administered online survey that was offered to people who were attempting to recover web-based assets using Warrick from the Internet Archive's Wayback Machine or search engine caches and follow-up in-depth interviews of survey-takers who were willing to submit to more extensive questioning. We consider these interviews supplementary, a prelude to more extensive qualitative research in the future.

The survey had 52 respondents, 34 of which were trying to recover a website that they had personally created, maintained, or owned, and 18 of which were trying to recover a website for someone else, a friend, relative, client, or in a few cases, for themselves to use as a resource; these responses were sufficiently complete to form a reliable picture of what happened. Eight responses too incomplete to warrant analysis were dropped.

The survey consisted of 75 questions, tailored to the two sub-populations of respondents (those recovering their own website vs. those recovering someone else's). Respondents recovering their own website answered 45 questions (11 of which were open-ended); respondents recovering someone else's website answered 33 questions (8 of which were open-ended). Respondents routinely skipped questions that they felt did not apply to them; for example, if they did not perform backups, they were unlikely to delve into the detailed coverage of how they did so.

The survey covered four basic areas: (1) a characterization of the website itself; (2) questions pertaining to the development and curation of the website, including where it was hosted and how it was backed up; (3) questions probing particular aspects of the loss and how it was discovered; and (4) questions about the restoration and how it did or did not influence the curation practices of the respondent. If the respondent was not responsible for the creation and curation of the original website, irrelevant questions were omitted and questions were rephrased to suit the circumstances.

Because the survey included so many open-ended questions, it proved to be useful in identifying directions to pursue during the interview portion of the study. Seven of the 52 respondents allowed us to interview them. The interviews were semi-structured and open-ended. Five of them took place over the phone and two international interviews took place using intensive Instant Messaging sessions over Skype. Interviews ran from a half hour to an hour and were recorded, and all voice recordings were transcribed.

To ground and focus the interviews, we asked preliminary questions that enabled us to look at the restored website whenever possible and center our questions around it; in one case, this was not possible, since the formerly public website was being recovered as a personal resource and was not destined for republication. Four of the interviewees were restoring websites they had created and maintained themselves; three were restoring websites they did not originally create. We also took the opportunity to ask more general questions about other digital belongings interviewees stored online. These questions will enable us to distinguish between website-specific curation practices and practices that pertain to digital belongings in general.

We also looked back on the data collected for a past study, described in [1], to extend the reach of the limited set of interviews conducted for this study. We isolated the portions of those 12 interviews that pertained to online material and used this data to triangulate the data gathered during our current study and to confirm or question the findings. Hence we had 19 sources of interview data as a window into general practices for curating online personal information.

## Respondents' Websites and Their Value

What kind of websites did survey respondents and interviewees think were sufficiently valuable to restore from caches and public archives? What made these websites valuable?

The websites described in the survey and discussed in the interviews spanned a spectrum of uses, from topical resources such as a Frank Sinatra fan site to web-based magazines to personal websites that respondents had created earlier in their lives (some quite extensive) to commercially important websites that advertised, provided information, and supported e-commerce for small businesses. Table 1 shows the breakdown of website genres, categorized by whether they were predominantly personal websites, had commercial value, were topical resources, were fan sites, were computer games, were publications, or were principally social venues; of course, this categorization is rough, and some of the websites spanned multiple genres.

**Table 1. Website purpose (by genre)**

| Type | Frequency (number) | Example |
|---|---|---|
| Topical | 29% (15) | Marijuana cultivation |
| Commercial | 19% (10) | Support for limousine business |
| Personal | 29% (15) | Art student's website |
| Fan site | 8% (4) | Frank Sinatra fan site |
| Publication | 8% (4) | Christian music e-zine |
| Social nexus | 8% (4) | Recent drama school graduates |

It should be no surprise that a significant proportion of the recovered websites had commercial value; what is more puzzling is why a commercially valuable website was lost to begin with. Three of the interviewees described commercial websites they were recovering; in all three cases, the web sites were not the main revenue source of the businesses they represented, yet they played a fundamental role. One supported the activities of a sports league (where the sports league itself was a revenue source for its two coordinators):

> "That's a big part of what we do. Just sort of enabling our players and our members to communicate with each other, be kept up-to-date in terms of what's going on with the league and games and stuff. So, I mean, the website is really a vital component of what we do."

Another advertised a house painting business in Florida:

> "It's good for a word of mouth and all I gotta do is say, 'go to [his name] dot com.' It's simple… It's on my business cards; it's on all my signs. And I've gotten people from, um, Ohio; I've gotten people from Chicago."

The third website was recovered for a client, a law firm that specializes in a particular type of product liability lawsuit:

> "They also did some work for a women's health alliance organization. They built a website years ago. And they did a lot of research and they had a lot of very specific drug fact information on there."

In each case, a different aspect of the website was considered valuable (besides the basic contact information); for the painter, it was the photos of his recently completed jobs; for the law firm, it was the extensive textual content, especially the transcription of a long speech; and for the sports league coordinator, it was the functionality and social nexus provided by the website.

The desire to recover topical resources and fan sites is also not particularly mysterious; these websites may supply unique and highly detailed information about an esoteric topic, information that may not be available anywhere else. The respondent who recovered the Frank Sinatra fan site told us: "First of all, I'm a nut for Frank Sinatra, so I'm interested in pretty much anything about him. But one of the things that really made me want to pursue snagging this data was that they had a lot of cross-reference stuff about his music and his albums." Thus she was recovering specific information that had immediate utility for her.

The recovery of personal websites is more difficult for respondents to explain since often what is there is of less concrete and immediate value. Without exception, personal websites were recovered by their original authors. One of the respondents who lost his personal website had difficulty characterizing why he was putting in so much effort to recover it since the website's value was largely emotional; he told us: "It was just data. You know, I didn't get my arms chopped off. Or get my heart broken or anything. It was just data." When asked if he'd want to listen to his personal podcasts again in 20 years, another respondent (who was interviewed via IM) wrote:

> "in 20 years? actually i don't listen to myself after i record. i usually record something and then check once or twice for dead air, edit and don't listen to myself again... i hate the sound of my voice."

Of course, some of the respondents who recovered personal websites have specific content in mind. One respondent told us over IM: "there's some pretty useful stuff in [the recovered site], if it's not outdated by 5 years already." He was referring to a series of rather technical 'how to' articles that he had written earlier.

In some cases, the recovered websites were fairly small (for example the house painter's website is around 5 Mbytes worth of job site photos); others were very extensive. Some of the personal and topical websites that made extensive use of in-line multimedia (as opposed to multimedia stored elsewhere) were in excess of a gigabyte. Complexity also varied greatly. Some websites were complex, interlinked sites that made extensive use of scripts and server-side processing and others were simple, composed using the basic content templates or web page editors provided by ISPs, and populated with text and photos.

If we think of a site's complexity (and therefore the likelihood that it can be reconstructed easily) as depending on whether it is dynamic and whether it has added functionality— some of it facilitating collaborative input— we find that well over a third of the sites were reasonably sophisticated. In fact, 21 out of the 52 respondents replied that they were recovering sites produced dynamically. Additional functionality varied too. Blogs were part of 21% of the lost websites, and forums were present in 31%. Interestingly, when asked specifically about this sort of facility, recovering personal blogs was considered important; other social content was adjudged to be ephemeral, especially given the difficulty of fully recovering it.

This distinction between important and ephemeral content often hinges its role. A respondent who recovered both his personal website and a commercial site, both with extensive blogs, said:

> "Actually, the Warrick tool, aside from restoring some of the stuff on the [commercial] blog, it was actually a little bit more important to me because it restored information from my personal blog. …The [commercial] blog wasn't the biggest thing in the world—to me, at least—but my personal blog goes all the way back to 2002 and is intensely important and personal."

It is impossible to predict whether a website is important by looking at the type or quantity of content or even by knowing its original purpose. Participants had a variety of reasons for recovering these websites including their emotional importance, the difficulty (or impossibility) of recreating the content, the time and cost involved in the original effort, the value of the information as a resource, an interest in reviving a community, and sometimes simply curiosity.

### *De facto* Archiving Strategies

Consumer strategies for keeping online digital material safe and archived for long-term access reflect a blend of opportunism, optimism, and benign neglect. We noticed three basic trends that arise from the characteristics of the current online environment and extend the way local digital belongings are handled:

- Materials are often opportunistically distributed over a variety of servers and services;
- Consumers employ circular reasoning about data safety; and
- Strategies based on benign neglect fail to take into account the server-side authoring capabilities offered by many current web hosting, blogging, and media sharing services.

We examine each trend in turn.

### *Distributing the files and spreading the risk*

First, consumers have learned to spread their risk and take advantage of the different free and low-cost storage services available on the Internet. Thus they might store photos on Flickr and videos on YouTube, create a blog on Blogger, publish a website on their ISP's server, and so on. Whether consciously or unconsciously, they realize that this mediates the risk of "losing everything" and provides them with functionality appropriate to the media type and their purposes. For example, an art student (specializing in animation) who has already lost several different portions of his personal webpage describes his strategy this way:

> "I keep backup lists because my site, blog, and podcast is currently on the free (for students here) website space our school generously provides. The problem is, I can't vouch for its permanence and so I set up backup lists for my peace of mind."

Thus he has reproduced partial copies of his website, his blog, his videos and animations, and his podcasts on different services.

Because each service has slightly varying capabilities, the copies are not necessarily equivalent. Some, as he notes, are better than others: one of his blog sites he has chosen because it allows him to have an easy-to-remember name; another he has chosen because he can partition the posts by subject. Remembering just where everything is and keeping all the mirrors up-to-date imposes a discernable tax on this strategy. It was not unusual during the interviews for a participant to suddenly recall a forgotten online store midway through our conversation: "I've posted some photos to, like, um, [pause] gosh I'm drawing a blank—oh! Pbase."

### *Circularity of reasoning: what protects what?*

Second, in part owing to this distribution of materials, respondents exhibit a pervasive circularity of reasoning about the safety of the files, databases, and code they rely on. First they might assert that even if the service or their account disappeared, they would still have the copy that they originally uploaded; then, in almost the same breath, they rationalize their home curatorial practices by saying that they would simply download the files from the web service they are using (never mind that they have reduced resolution or otherwise culled material to post it online). For example, one respondent told us he did not worry unduly about his valuable photos:

> "The good thing about the photos is that there's always an intermediary step. I mean like the photos go off of my camera onto my computer before they go up to Flickr. So I always have master copies on my PC. So that's why I don't care so much about Flickr evaporating."

Not long after making this claim, he said, "But in the event of a catastrophe, presumably I could log into Flickr and be like, 'Hey. Send me a DVD. Stat.'" This circularity is not unusual; in fact we noticed it to some degree in all of our interviews.

Several respondents who used circular reasoning like this did note that their circular safety net resulted in imperfect recovery. One interviewee who had lost online photos due to account inactivity told us:

> I didn't lose the pictures, but I was sorry that I had lost the collections and the organization and, you know. I'm sure I have the pictures somewhere still. But fishing them out and recreating it was not feasible.

Thus, while much of the content is indeed copied, the copies are not necessarily complete nor exact. Furthermore, as we suggest in the next subsection, additional material is often added on the server side and not retained locally.

### *Benign neglect and server-side augmentation*

Finally, as we saw with local digital belongings in a previous study [1], there was considerable evidence of benign neglect, an adoption of an almost value-neutral stance in which loss has been rationalized in advance. For example, many of our respondents used server-side tools for constructing their websites; these tools would use simple dynamic capabilities to knit together a website from the uploaded materials and elicit additional description from the author. Although interviewees realized that they would lose descriptions and structure that they had added during the server-side website construction process, they were unwilling to put forth any curatorial effort to ensure they would not lose this work.

One participant who had recovered his personal blog noted:

> "There're literally hundreds of posts. And not to mention the fact that I wouldn't even necessarily have a perfect memory of, like, y'know, whether a post existed or not. So even if I did look through every single post, it's not possible for me to really know for sure if one got missed or not. Because I just wouldn't remember myself. You just had to trust it. I had to run the batch process, do some spot checks, and if it looked good, then I just kind of had to resign myself to the fact that that's just what it was. If there were posts that were missed, then that's just the price I paid for not backing up."

But these websites represent material that is crawled and cached by a number of different public stores. What of other types of web-based personal material such as email? Even if they are distinctly valuable, respondents seem to give little thought to their long-term safety. One participant said:

> "I have a gmail account that is probably even more important than … my website. If I lost my gmail account and all my associated email, I'd probably have a schizophrenic episode or something. Because I use it for more than email. I email myself just important little chunks of data… online email…makes it convenient for throwing files up in a sort of protected way."

After some thought, he realized that because he used POP, he had a second copy of these important files, but there was scant evidence that he felt he should expend any extra effort to ensure that these files were archived. In fact, he described this way of thinking as "quaint."

## How Websites Are Lost

We often make assumptions about how digital files are lost: for example, through hardware malfunctions, through media deterioration, through accidental deletion, through format obsolescence, or through file corruption. These assumptions lead us to propose certain solutions for keeping archival material safe: for example, implementing a backup regimen to protect against hardware malfunctions, refreshing media to ensure that media deterioration is kept at bay, storing valuable files in standardized formats, and making additional copies of important files so that they can be replaced in the event of accidental deletion or corrupting edits.

So how do people lose material they publish on the web? Our past studies have revealed that, to a great degree, people trust Internet Service Providers (ISP) or companies that provide specific services such as Flickr or Yahoo Mail to keep their online files safe – to implement good IT practices including backups and routine maintenance and to notify their customers if there is a problem (including the service provider discontinuing the service or going out of business) [1]. In other words, they take no extra precautions themselves and might even be surprised if they were told there was a need to do so.

Furthermore, adhering to a principal of benign neglect, individuals sometimes forget they even have files that might disappear with an unused account – transitions such as graduation or loss of a job are not necessarily triggers for moving files; furthermore, some free services have policies that cause accounts to be deleted when they have been dormant for awhile.

More than half of our interviewees were recovering websites lost in this way. Some are still unsure of why their accounts disappeared. One respondent told us via IM: "geocities deleted the entire account for a reason unknown to me; it was inactive for a while, maybe. I wasn't able to find out [why]. could be that the account was registered to an email address that was dropped and they sent a notice there." Another respondent explained (over IM) how his podcasts disappeared:

> "i hosted my podcasts early on on a free service called Rizzn.net… he then changed rizzn.net to something called blipmedia.com and then!! he decided to sell blipmedia … and he never emailed people about it.. suddenly the files were gone and the only news i heard about it was when i had to hunt online for what happened… and in blipmedia's google help group it was only when people ASKED HIM ABOUT IT that he explained."

It is remarkable that our interviewees not only cared enough to spend considerable effort recovering their lost websites, but also that many continue to look for creative ways to replace lost files. One interviewee sent us a five year old ftp log and some HTML template files to enlist additional assistance in recovering his files. Yet despite the acknowledged personal, informational, or commercial value of the materials, they were still overseen with a large degree of benign neglect.

Nor do individuals necessarily plan for circumstances like police raids or hacking: they tend to attribute file loss to some sort of technological mishap, not to maliciousness or illegal activity. One respondent complained that his or her site was lost when another, presumably illegal, site on the host's computers was confiscated by the police. Most ISPs have no provision in the event of death of the account owner; one respondent said his Web files were lost because "[The] hosting server [was] taken off-line without notice when the author died. [There was] no contact information for retrieval of data and no next of kin to obtain backups from."

Table 2 shows a breakdown of general causes of Web site loss as reported by survey respondents. It is interesting to note that many of the losses are not associated with a catastrophic technological event; rather they represent simple neglect – an account was forgotten and subsequently lost or a trusted service was discontinued. Furthermore, even the more stereotypical losses stem from simple misplacement of trust – users trusted their ISPs to back up files and their ISPs did not do so. Some of the losses reveal that individuals are not aware of or do not understand their ISPs' policies. For example, one respondent said, "Geocities deleted the website without warning. Possible cause could be presence of several pages that described hacking Kazaa to remove unwanted content/bundled programs."

**Table 2. Reasons for Website Loss**

| Type | Frequency (number) |
|---|---|
| Service/server discontinued | 33% (17) |
| ISP IT policies and practices | 19% (10) |
| Unknown | 13% (7) |
| Hacking | 10% (5) |
| Lost account | 8% (4) |
| Hard drive failure | 6% (3) |
| Owner deletion | 6% (3) |
| Police raid | 4% (2) |
| Death | 2% (1) |

This breakdown demonstrates the wisdom of developing methods of personal digital archiving that anticipate the benign neglect of distributed and augmented materials that we described in the previous section. Many individuals are unaware of the specific IT practices of their ISPs (for example, how regularly their files are backed up or whether they are backed up at all); nor do they keep careful track of the status of their various accounts or the ISPs policies regarding account dormancy. In fact, the survey responses indicate that the respondents regard their websites as archival or permanent, and the service providers do not.

## Discovery of Loss

When personal digital assets are stored locally, we usually assume that we play an active role in preventing and discovering their loss, even if in practice we do little to implement hedges against this loss. In the best case, we might try to intervene – to copy files to another storage medium or to perform preventive maintenance more often – when, for example, we hear our local computer's disk drive making funny noises. The subsequent data loss is no real surprise and its discovery is more or less immediate.

However, our respondents demonstrated that this assumption does not necessarily hold for online digital assets. Results of the survey and the interviews reveal three important distinctions between locally stored material and Internet server-based digital assets:

- There is a mismatch between an owner's expectation of asset value and their ISP's notification policies and procedures;
- There is often a greater temporal gap between the site's disappearance, detection of the loss, and recovery of the material than we would expect; and
- There is often a discrepancy between site owner's perception of the permanence of online materials and the actual *ad hoc* nature of many network services.

Instead of learning that their websites or other digital holdings were intentionally deleted, removed according to policy, or lost in a technology snafu, many of our respondents discovered the loss more indirectly. Often they realized something was amiss when they tried to access the website themselves, or when friends, colleagues, customers, or clients went looking for the material and told them it was gone. For example, when one respondent tried to access his or her website, "the URL was no longer valid." In a few cases, maintenance or update efforts made respondents aware of the loss. Sometimes use of a missing subsystem on the website – email or IRC – was the red flag. In one case, one in which the web server was confiscated in a police raid, the respondent heard about his missing website in news reports.

It is less common than we would expect for notification to be active and persistent, for the ISP or the server owner to establish contact with the customer to tell him or her that the material is in jeopardy or gone. In most cases, it seems as if the ISP is assuming that the customer's files are worthless unless someone says something to the contrary. One survey respondent told us that he or she noticed something was wrong because "the data isn't up to date, the data was [sic] came from backup." Another respondent wrote: "The website was replaced with a single page explaining the circumstances under which it was lost." If the site has been removed for reasons of policy, the notification may be more active, as it was in the case of this suspected violation: "several days ago I was directed to obtain rights to reproduce content."

In fact, when respondents were asked how long the website in question – sites they had created themselves – had been gone before the loss was discovered, a significant proportion did not notice the loss immediately, suggesting that this was not a highly active site and that the website's owner had an expectation of permanence. A respondent who recovered a six year old website for a client explained:

> "They did a lot of research and they had a lot of very specific drug fact information on there. And then they built it and had someone hosting it for them. And then that person, they couldn't contact anymore. They wanted to make changes, and then the website went down, and they couldn't find him anymore. So he just kind of disappeared."

Not only did the website's owners expect the site to be more or less permanent, they were also unaware how *ad hoc* the website hosting situation was.

Temporal discrepancies are even more pronounced in sites recovered by someone other than its owner. Site owners in our survey usually noticed the site's disappearance in under a week (almost 65% did) and began to substantially restore it in under a week (about 45%); but over 40% of non-owners waited more than a year (sometimes significantly more than a year) after the site disappeared to restore it. This lapse suggests that the digital assets do have long-term value in spite of the incaution on the part of the sites' owners. In fact, the respondent who was recovering the Sinatra fan site told us:

> "Somewhere they had a link to this [Sinatra fan] site. 'You'll find a lot of great info'… 'at thus-and-so web address.' And I tried to go there. And of course it said, not found or whatever…. So I'd look every week or something to see if it was back up, just assuming that it was down temporarily. Then after a few weeks, it seemed to me that it was not coming back and if I wanted to get that information, I would have to somehow get it from these [search engine] caches."

She had never seen the original site; yet she had ascertained that it was of sufficient value to warrant her recovery efforts.

## Conclusion

Curation of personal digital materials in online stores bears some striking similarities to the curation of similar materials stored locally. Participants continue to archive personal assets by relying on a combination of benign neglect, sporadic backups, and unsystematic file replication. However, the story does not stop there. We have identified some problems specific to online stores, ways in which expectations of permanence and notification are violated, and situations in which benign neglect has far greater consequences than it does for local materials.

Before they lost their websites, between a quarter and a third of the respondents who maintained their own websites maintained no backup at all, let alone a formal archive of valuable digital assets. In a few cases, this loss was a wake-up call that provoked respondents to consider instituting some sort of backup procedure in the future (however at this writing, even 6 months after the survey, these good intentions have not been realized); but in other cases, the respondents simply retrenched after recovering some or all of their lost material. They believed that the problem was solved by simply switching service providers. A few respondents interpreted their satisfactory file recovery results using Warrick as an invitation to avoid the extra legwork of maintaining their own archive: "Google has it all anyway. And their backup is much more reliable than mine could ever hope to be."

It is difficult to believe that any of these common practices – *ad hoc* use of local storage as file backup, incidental use of felicitous institutional or corporate caches or archives, or implicit reliance on an ISP's good IT practices – are oriented toward long term archiving or that they will produce satisfactory results twenty, fifty, or a hundred years hence. Distributed storage places a burden on the individual to keep track of where everything is and its status. Consumers and small business owners need to be aware of ISPs' varying standards, policies, and practices, some of which are surely not in the negligent consumer's best interest. This study has demonstrated the variability of factors central to the long-term health of valuable files such as notification of significant account changes, data retention policies connected with account deactivation, the irregular implementation of routine IT services like backup, not to mention the curatorial practices that would enable material to be viable long into the future.

Then too, maintaining an archival copy of material stored in many of these online services is a complex matter. In the so-called Web 2.0 environment, much of the value relies on server-side social activity and additional computation; even in the simpler case of dynamic web pages and shared repositories, significant structure and descriptive metadata is added after-the-fact. The downfall of the circularity of many of our participants' long term storage strategies becomes all too evident.

In the end, it is likely that no single set of best practices will work for all individuals nor will a single archiving technology address all of the complex problems of maintaining one's personal digital belongings over a lifetime and beyond. This study has enabled us to evaluate the likelihood that Internet-based digital belongings will survive over extended periods and assess nature of the loss associated with personal files stored on Internet-based services; it has also given us some insight into the viability of lazy preservation (as a technological strategy compatible with benign neglect) in the larger sphere of personal digital archiving.

## References

[1] Marshall, C.C., Bly, S., and Brun-Cottan, F., The Long Term Fate of Our Digital Belongings, *Proc. IS&T Archiving '06*, pg. 25-30. (2006).
[2] Marshall, C.C. How People Manage Personal Information over a Lifetime. In *Personal Information Management: Challenges and Opportunities* (UW Press, Seattle, WA, 2007), pg. 85.
[3] McCown, F., Smith, J.A., Nelson, M.L, and Bollen, J., Lazy Preservation: Reconstructing Websites by Crawling the Crawlers. *Proc. ACM WIDM '06*, pg. 67-74. (2006).
[4] Smith, J.A., McCown, F., and Nelson, M.L., Observed Web Robot Behavior on Decaying Web Subsites, *D-Lib Magazine*, 12, 2. (2006).
[5] Wayback Machine, http://www.archive.org/web/web.php.

## Author Biography

Cathy Marshall is a Senior Researcher at Microsoft; for more information, see her homepage at http://www.csdl.tamu.edu/~marshall.
Frank McCown is a Ph.D. candidate at Old Dominion University (ODU).
Michael Nelson is an assistant professor of computer science at ODU.